\documentclass[12pt]{article}
\usepackage{jheppub}
\usepackage{amssymb,amsmath,amstext,amsfonts}
\usepackage{bbold,ulem,bm}
\usepackage{graphics}
\usepackage{mathtools}
\usepackage{hyperref}
\usepackage{color}

\title{First Sound in Holographic Superfluids at Zero Temperature} 

\author{Angelo Esposito,}
\author{Sebastian Garcia-Saenz,}
\author{and Riccardo Penco}

\affiliation{Physics Department, Center for Theoretical Physics \\
\& Institute for Strings, Cosmology, and Astroparticle Physics,\\
  Columbia University, New York, NY 10027, USA}

\abstract{Within the context of AdS/CFT, the gravity dual of an s-wave superfluid is given by scalar QED on an asymptotically AdS spacetime. While this conclusion is vastly supported by numerical arguments, here we provide an analytical proof that this is indeed the case. Working at zero temperature, we explicitly find the quadratic action for the superfluid phonon at the boundary in an arbitrary number of dimensions and for an arbitrary scalar field potential, recovering the known dispersion relation for conformal first sound.}

\begin{document}

\maketitle
\flushbottom
\thispagestyle{empty}
\newpage
\setcounter{page}{1}

\section{Introduction}

In recent years the AdS/CFT correspondence~\cite{Maldacena:1997re,Gubser:1998bc,Witten:1998qj} has found numerous applications in the study of condensed matter systems---see~\cite{Hartnoll:2009sz,Herzog:2009xv} for a review and~\cite{Zaanen:2015oix} for a more recent textbook treatment. The correspondence, in fact, provides a powerful tool to study strongly coupled systems using well-established high energy theory techniques. Such strongly coupled systems not only are ubiquitous in nature but they can also be engineered and studied in the laboratory.
 
Particularly interesting is the use of such gauge/gravity duality to study systems in a phase of spontaneously broken symmetry. Arguably one of the simplest systems of this sort is the zero temperature s-wave superfluid, where a spontaneously broken $U(1)$ charge is at finite density~\cite{Nicolis:2011pv}. At low energies, such system is described by a single real scalar field, $\phi$, that shifts under $U(1)$ and acquires a time-dependent expectation value $\langle\phi\rangle=\mu t$, with $\mu$ the chemical potential for the $U(1)$ charge. This expectation value spontaneously breaks boosts, as well as $U(1)$ and time translations down to the diagonal subgroup~\cite{Nicolis:2013lma}. Nevertheless, this system admits a single Goldstone boson---the phonon---associated with fluctuations of $\phi$ around its background, $\phi=\mu t+\pi$. In the relativistic case, the low-energy action for phonons can be written as~\cite{Son:2002zn}
\begin{align} \label{eq:X}
S=\int d^Dx\,P(X),\;\;\text{ with }\;\;X=-\partial_\mu\phi\partial^\mu\phi.
\end{align}
The functional form of $P(X)$ determines (implicitly) the superfluid equation of state. 

The same model of superfluidity can also be derived by interpreting $\phi$ as the phase of a complex field $\Phi$ charged under $U(1)$, with time derivatives shifted by the chemical potential:
\begin{align}
\Phi(x)=e^{i\phi(x)}, \quad\quad \partial_t\to\partial_t-i\mu.
\end{align}
In this language $\langle\phi\rangle=0$ and $\mu$ can be thought of as the expectation value of the temporal component of a gauge field. It should be easy to convince oneself that these two viewpoints are completely equivalent, albeit the latter one is perhaps more common in the holographic literature.

From Eq.\ (\ref{eq:X}), we obtain the general expression for the stress-energy tensor of a superfluid at zero temperature:
\begin{align}
T_{\mu\nu}=-\frac{2}{\sqrt{-g}}\frac{\delta S}{\delta g^{\mu\nu}}=2P_X\partial_\mu\phi\partial_\nu\phi+\eta_{\mu\nu}P,
\end{align}
where $P_X$ stands for the derivative of $P$ with respect to $X$. In this paper we will be particularly interested in conformal superfluids. In this case, the traceless condition $T_{\;\;\mu}^\mu=0$ completely fixes the action up to an overall normalization factor: $P(X)\propto X^{D/2}$. This also determines the phonon sound speed, which can be shown to be~\cite{Nicolis:2011cs}
\begin{align} \label{eq:cs}
c_s^2=\frac{P_X}{P_X+2XP_{XX}}=\frac{1}{D-1}.
\end{align}

Conformal s-wave superfluids admit a simple dual gravitational description: scalar QED on an asymptotically AdS spacetime~\cite{Herzog:2008he}. Based on the standard holographic dictionary, the $U(1)$ gauge symmetry in the bulk of AdS is the counterpart of the global $U(1)$ symmetry on the boundary. The latter is spontaneously broken because the charged boundary operator dual to the bulk scalar field acquires a vacuum expectation value (VEV), and it is at finite density because the VEV of the $t$-component of the gauge field acts as a source for the charge density. These are indeed the defining features of a superfluid. 

Note that scalar QED can also be regarded as a model for superconductivity~\cite{Hartnoll:2008vx,Hartnoll:2008kx} by virtue of the fact that the charge response of a superconductor is described by superfluid hydrodynamics. However, when it comes to the low-energy spectrum of excitations the distinction between superfluids and superconductors is important: superfluids have a gapless Goldstone mode (the phonon), whereas superconductors do not (since the would-be Goldstone is ``eaten'' via the usual Higgs mechanism). As we will see, the boundary theory that is dual to an abelian Higgs model in the bulk has in fact a gapless mode.

The identification of scalar QED with a superfluid on the boundary has so far---to the best of our knowledge---been established based on results that are partly numerical. Specifically, studies of first, second and fourth sound in holographic superfluids have been performed in~\cite{Herzog:2008he,Herzog:2009md,Yarom:2009uq}. In the present paper we would like to further address this by providing an analytic derivation of the quadratic effective action at the boundary. To this end, we will first use a simple scaling argument, valid on a fixed AdS background, to show how the background boundary action depends on the chemical potential $\mu$. Next we will employ similar techniques to those developed in~\cite{Nickel:2010pr,deBoer:2015ija,Crossley:2015tka,Argurio:2015via,Argurio:2015wgr} to derive the quadratic action for the phonons on the boundary. We will show in particular that one recovers the correct dispersion relation for the first sound of a conformal superfluid.

\vspace{1em}

\noindent {\it Conventions:} Throughout this paper, we will work in units such that $\hbar = c = L =1$, (where $L$ is the AdS radius) and we will adopt a ``mostly plus'' metric signature. $D$ is the spacetime dimension of the boundary theory. We will use capital latin letters $M, N, ...$ for the bulk indices, which are contracted with the AdS metric, and greek indices $\mu, \nu, ... $ for the boundary indices, which are contracted with the Minkowski metric.\\

\section{Background boundary action}

\subsection{Set-up}

The gravity dual of a superfluid is described by the following Maxwell-scalar field action:
\begin{align} \label{eq:S}
S &= - \int d^{D+1}x\,\sqrt{- g}\bigg[\left|\partial \Phi-iqA\Phi\right|^2+V\!\left(|\Phi|^2\right)+\frac{1}{4}F_{MN}F^{MN}\bigg]+S_\text{c.t.}.
\end{align} 
Without a string theory embedding there are no constraints on the form of the potential $V$, and as such it should be regarded as a phenomenological quantity. Throughout the paper we will keep it completely general, assuming that it is of the form
\begin{align}
V\!\left(|\Phi|^2\right)=m^2{|\Phi|}^2+\text{interaction terms}.
\end{align}
Since we wish to work at zero temperature, we will start with a pure AdS$_{D+1}$ metric in Poincar\'{e} coordinates:
\begin{align} \label{ds2}
ds^2=\frac{d x^\mu dx_\mu+du^2}{u^2}.
\end{align}
In these coordinates the AdS boundary is located at \mbox{$u=0$} while the horizon is at $u=\infty$. In the rest of the paper we will neglect the gravitational backreaction and take the AdS background as fixed. Neglecting the backreaction of the fields on the spacetime geometry can be formally achieved by taking the charge $q$ to infinity, the so-called ``probe limit''~\cite{Yarom:2009uq}.

It is important to remark, however, that the metric~\eqref{ds2} does not always provide the correct description for the geometry of the ground state holographic superfluid. For many different choices of the potential $V(|\Phi|^2)$ the metric does not exhibit conformal symmetry in the infrared~\cite{Horowitz:2009ij,Gubser:2009cg}, and hence our approximation would be incorrect.\footnote{We thank C.\ Herzog and A.\ Yarom for bringing this point and the relevant references to our attention.} Nevertheless, there exist specific theories for which the background fields stay finite and the metric is AdS both near $u=0$ and near $u=\infty$. This is in fact what occurs, for example, for a free massless scalar field~\cite{Horowitz:2009ij}, or for a W-shaped potential of the kind~\cite{Gubser:2008wz,Gubser:2009cg}
\begin{align} \label{Wpot}
V\!\left(|\Phi|^2\right)=m^2{|\Phi|}^2+\frac{\lambda}{2}{|\Phi|}^4,
\end{align}
assuming $m^2<0$, $\lambda>0$ and large charge $q$. For these theories our ansatz~\eqref{ds2} and the non-backreaction approximation are consistent, and so we will assume from now on that we are working with such potentials.

The plane symmetric ansatz for the background fields is
\begin{equation} \label{ansatz}
\Phi\equiv \rho(u),\qquad\quad A_M\equiv\sqrt{2}\,\frac{\psi(u)}{u}\delta_M^0,
\end{equation}
and the expression for the gauge field is chosen for later convenience. The near-boundary behavior of the fields is
\begin{subequations} \label{falloff}
\begin{align}
\rho&=\rho_{(1)}u^{D-\Delta}+\rho_{(2)}u^{\Delta}+\cdots, \label{falloffs} \\
\psi&=\frac{\mu}{\sqrt{2}}u - \frac{\varepsilon}{\sqrt{2}}u^{D-1}+\cdots, \label{asymppsi}
\end{align}
\end{subequations}
where $\Delta>0$ is the larger of the two solutions of $\Delta(\Delta-D)=m^2$. The falloffs of the gauge field $\mu$ and $\varepsilon$ are respectively the chemical potential and the charge density of the dual theory.

From now on we will impose the $\rho_{(1)}=0$ boundary condition, which ensures that the background field is normalizable in the case $m^2\geq -D^2/4+1$. In addition, to fix the variational problem we further need to include a counterterm action as in (\ref{eq:S}). A general expression for such action in an arbitrary number of dimensions is not available. Nevertheless, as we will show in Sec.~\ref{phonons}, its detailed knowledge is not needed for our analysis.

Notice that the choice $\rho_{(1)}=0$ also implies that the boundary operator dual to the scalar field have a nonzero VEV but no source, so that the $U(1)$ symmetry is broken spontaneously.

\subsection{On-shell action}

The equations of motion for the background fields read
\begin{subequations} \label{eq:bkg}
\begin{gather}
\rho^{\prime\prime}-\frac{D-1}{u}\rho^\prime-\frac{1}{u^2}V^\prime\!\left(\rho^2\right)\rho+2q^2\frac{\psi^2}{u^2}\rho=0, \label{eq:bkg1} \\
\psi^{\prime\prime}-\frac{D-1}{u}\psi^\prime+\frac{D-1}{u^2}\psi-2q^2\frac{\rho^2}{u^2}\psi=0, \label{eq:bkg2}
\end{gather}
\end{subequations}
where primes on the fields represent derivatives with respect to the radial coordinate $u$. They should not be confused with $V^\prime$, the derivative of the potential with respect to its argument.

Let us now consider a particular solution to the previous equations with $\mu=1$ and let us denote such fields as $\hat\rho(u)$ and $\hat\psi(u)$. Since the equations~\eqref{eq:bkg} are invariant under rescaling of the $u$ coordinate, it follows that the general solution with chemical potential $\mu$ can be obtained from the one with $\mu =1$ by replacing $u\to\mu u$. In other words, we have $\rho(u)=\hat\rho(\mu u)$ and $\psi(u)=\hat\psi(\mu u)$.

With this in mind we can now calculate how the boundary action evaluated on the background depends on the chemical potential $\mu$. The background on-shell action is given by
\begin{align}
S_{\text{bkg}}&= - \int\frac{du\,d^Dx}{u^{D+1}}\Bigg[u^2(\rho^\prime)^2+V\!\left(\rho^2\right)+2q^2\psi^2\rho^2+\frac{u^4}{2}\left(\frac{\psi^\prime}{u}-\frac{\psi}{u^2}\right)^2\bigg] \equiv\mathcal{N}\int d^Dx \, \mu^D, \label{eq:muD}
\end{align}
where, in the last step, we have performed the integral over $u$ by introducing the rescaled variable $y=\mu u$, and we have defined the overall constant
\begin{align}
\mathcal{N}&\equiv- \int \frac{dy}{y^{D+1}}\Bigg[y^2(\hat\rho^\prime)^2+V\!\left(\hat\rho^2\right)-2q^2\hat\psi^2\hat\rho^2-\frac{y^4}{2}\left(\frac{\hat\psi^\prime}{y}-\frac{\hat\psi}{y^2}\right)^2\Bigg].
\end{align}
It is important to stress that the chemical potential $\mu$ does not appear anywhere in the definition of the constant $\mathcal N$. Therefore, the bulk Lagrangian evaluated on the background scales like $\mu^D$, which suggests the $X^{D/2}$ behavior of a conformal superfluid---see Eq.~\eqref{eq:X}. We remark, however, that this is a consequence of simple dimensional analysis and does not provide by itself a proof that we are indeed dealing with a superfluid, since it does not give the dependence of the action on the phonon field. To this end we now turn to study perturbations around the background configurations (\ref{ansatz}).\\

\section{Quadratic action for phonons} \label{phonons}

Although an analytic solution to Eqs.\ \eqref{eq:bkg} is not available, we will now show that the low energy effective action for the boundary Goldstone modes can nevertheless be calculated explicitly.

In order to do that, we introduce the fluctuations of the scalar and gauge field via
\begin{align}
\Phi=(\rho+\sigma)e^{i\pi},\quad\quad A_M=\bar A_M+\alpha_M,
\end{align}
where we defined $\bar A_M=(\sqrt{2}\,\psi/u)\delta_M^0$. The quadratic bulk action for the fluctuations then reads:
\begin{align} \label{eq:S2}
S^{(2)}=&-\int d^{D+1}x\,\sqrt{-g}\bigg[\partial_M\sigma\partial^M\sigma+\rho^2\partial_M\pi\partial^M\pi-4q\bar A^M\rho\,\partial_M\pi\,\sigma \notag \\
&-2q\rho^2\alpha^M\partial_M\pi+q^2\bar A^M\bar A_M\sigma^2+4q^2\rho\bar A_M\alpha^M\sigma+q^2\rho^2\alpha_M\alpha^M \notag \\
&+\left(V^\prime+2\rho^2V^{\prime\prime}\right)\sigma^2+\frac{1}{4}f_{MN} f^{MN}\bigg] +S^{(2)}_\text{c.t.}\,,
\end{align}
where $f_{MN}=\partial_M\alpha_N-\partial_N\alpha_M$ and $S_\text{c.t.}^{(2)}$ is the part of the counterterm action quadratic in the fluctuations. From now on $V\!\equiv \!V\!\left(\rho^2\right)$ is the potential computed on the background. The plan of action is now similar to the one presented in \cite{deBoer:2015ija}: we will solve the linearized equations of motion for all the fluctuations but $\pi$ to lowest order in boundary derivatives, and then plug the solutions back into the original action to obtain the action for the Goldstone bosons at the boundary. We will be able to carry out this procedure without fixing any gauge~\cite{Nickel:2010pr,deBoer:2015ija}.

\vspace{1em}

Since the phase $\pi$ always appears with a boundary derivative (except for Eq.\ \eqref{eq:pi} below, which plays a special role), to implement the low energy expansion of the bulk theory we will assume the following scaling rules~\cite{deBoer:2015ija}:
\begin{eqnarray}
\sigma, \, \alpha_\mu, \,  \partial_u \sim \mathcal{O} (1), \qquad  \partial_\mu \sim \mathcal{O} (\epsilon), \qquad \pi, \, \alpha_u \sim \mathcal{O} (1/\epsilon), \nonumber 
\end{eqnarray}
where the scaling of $\alpha_u$ follows from consistency with the equations of motion, which at lowest order in $\epsilon$ are 
\begin{subequations} \label{eq:eom}
\begin{gather}
\sigma^{\prime\prime}-\frac{(D-1)}{u}\,\sigma^\prime-\frac{1}{u^2}\left( V^\prime+2\rho^2V^{\prime\prime} \right)\sigma+2\sqrt{2}\,\frac{q\rho\psi}{u}(q\alpha_0-\partial_0{\pi})+\frac{2q^2\psi^2}{u^2}\,\sigma=0\,, \label{eq:sigma} \\
{(q\alpha_0-\partial_0\pi)}^{\prime\prime}-\frac{(D-3)}{u}{(q\alpha_0-\partial_0\pi)}^{\prime}-2\,\frac{q^2\rho^2}{u^2}(q\alpha_0-\partial_0\pi)-4\sqrt{2}\,\frac{q^3\rho\psi}{u^3}\,\sigma=0\,, \label{eq:at} \\
{(q\alpha_i-\partial_i\pi)}^{\prime\prime}-\frac{(D-3)}{u}{(q\alpha_i-\partial_i\pi)}^{\prime}-2\,\frac{q^2\rho^2}{u^2}(q\alpha_i-\partial_i\pi)=0\,, \label{eq:ai} \\
\pi^\prime-q\alpha_u=0\,. \label{eq:pi}
\end{gather}
\end{subequations}
The equations for $\sigma$ and the gauge-invariant combination $q\alpha_\mu-\partial_\mu\pi$ are of second order, but the one for $\pi$ is only of first order. As a result, we need two boundary conditions for $\sigma$ and $\alpha_\mu$, but only one for $\pi$. We will choose Dirichlet boundary conditions for all the fields. In particular, we require for $\pi$ to vanish at $u=\infty$,\footnote{This is the correct boundary condition for $\pi$ only if we do not fix any gauge. If for instance we worked in a gauge where $\alpha_u = 0$, then we would need to impose $\pi = \pi_B (x^\mu)$ at $u = \infty$~\cite{Nickel:2010pr,deBoer:2015ija}.} so that Eq.~\eqref{eq:pi} immediately gives
\begin{equation} \label{eq:pi_sol}
\pi=-q\int^\infty_u dw\,\alpha_u(w)\,.
\end{equation}
We will see below that the Goldstone field is identified with the boundary value of the scalar $\pi$, i.e.\ $\pi_B(x^{\mu})\equiv\pi(u=0,x^{\mu})$. This is why we will use all the equations of motion but the one for $\pi$. On the other hand we will impose for $\sigma$ and $\alpha_\mu$ to vanish both at $u=0$ and $u=\infty$.

Before proceeding, though, it is worth addressing at this point a subtlety hidden in Eqs.\ \eqref{eq:eom}. The low energy expansion is not, strictly speaking, appropriate for every value of $u$, since in a region close to the center of the AdS spacetime (large $u$) the $\mathcal{O}(\epsilon^2)$ terms cannot be neglected anymore. This poses an issue regarding the order in which we choose to take the two limits: the low energy limit and the $u\to\infty$ limit. The rigorous way to treat this problem would be to first introduce an IR cutoff at some large $u=\Lambda$. For each value of the cutoff it is safe to take the low energy limit, solve the equations of motion and find the boundary action. Then, at the very end, one should remove the cutoff by sending $\Lambda\to\infty$. This procedure is rather long and cumbersome and we will therefore be slightly cavalier about it. It can be checked however that the final result is not affected by which limit is taken first.

To solve the remaining equations in~\eqref{eq:eom}, consider first the change of variable
\begin{align}
q\alpha_i-\partial_i\pi\equiv \frac{\beta_i(u)}{u}\,.
\end{align}
Substituting this in \eqref{eq:ai} yields
\begin{align}
\beta_i^{\prime\prime}-\frac{D-1}{u}\,\beta_i^\prime+\frac{D-1}{u^2}\,\beta_i-2\,\frac{q^2\rho^2}{u^2}\,\beta_i=0\,,
\end{align}
which is precisely the equation satisfied by the background function $\psi(u)$ (see \eqref{eq:bkg2}), and therefore a regular solution is simply $\beta_i=b_i(x^{\mu})\psi(u)$. By imposing vanishing Dirichlet boundary conditions on $\alpha_i$ we can determine $b_i(x^{\mu})$ and finally arrive at
\begin{align} \label{eq:ai_sol}
\alpha_i=\frac{1}{q}\left(\partial_i\pi-\frac{\sqrt{2}\,\psi}{\mu u}\,\partial_i\pi_B\right)\,.
\end{align}

We can similarly solve for $\sigma$ and $\alpha_0$ in terms of $\psi$ and $\rho$ by defining
\begin{align}
\sigma\equiv u\gamma'(u)\,,\qquad q\alpha_0-\partial_0\pi\equiv q\sqrt{2}\,\delta'(u)\,.
\end{align}
From Eqs. \eqref{eq:sigma} and \eqref{eq:at} we then obtain
\begin{subequations} \label{gamma delta eqs}
\begin{alignat}{3}
\gamma'''-\frac{(D-3)}{u}\,\gamma''-\frac{\left( D-1+V'+2\rho^2V'' \right)}{u^2}\,\gamma'+4\,\frac{q^2\rho\psi}{u^2}\,\delta'+2\,\frac{q^2\psi^2}{u^2}\,\gamma'=0\,,&\\
\delta'''-\frac{(D-3)}{u}\,\delta''-4\,\frac{q^2\rho\psi}{u^2}\,\gamma'-2\,\frac{q^2\rho^2}{u^2}\,\delta'=0\,.&
\end{alignat}
\end{subequations}
If we replace $\gamma, \delta$ in these equations with $\rho, \psi$ respectively, we recover the derivative of the background equations \eqref{eq:bkg}. This suggests that regular solutions to Eqs.\ (\ref{gamma delta eqs}) are given by $\gamma=c(x^{\mu})\rho(u)$ and $\delta=c(x^{\mu})\psi(u)$ (up to an irrelevant integration function independent of $u$). Notice incidentally that the proportionality function $c$ must be the same in $\gamma$ and $\delta$, as one can verify using the above equations. Imposing again vanishing Dirichlet boundary conditions for $\sigma$ and $\alpha_0$ produces the results
\begin{align} \label{eq:sigma_a0_sol}
\sigma=-\frac{\rho'}{q\mu}\,u\partial_0\pi_B,\;\;\; \alpha_0=\frac{1}{q}\left(\partial_0\pi-\frac{\sqrt{2}\,\psi'}{\mu}\,\partial_0\pi_B\right)\,.
\end{align}
These solutions correctly vanish in the IR limit if the background fields $\psi$ and $\rho$ are regular, i.e.\ if they behave as $\psi,\rho\sim {\mathrm{constant}} +u^{-\nu}$ as $u\to\infty$, with $\nu$ large enough. For example, this is what was found in~\cite{Horowitz:2009ij} for the case of a free massless field and in~\cite{Gubser:2008wz} for the potential~\eqref{Wpot}.

\vspace{1em}

Eqs.~\eqref{eq:ai_sol} and \eqref{eq:sigma_a0_sol} contain the solutions for the perturbation fields $\alpha_{\mu}$ and $\sigma$ in terms of $\pi$ and the fixed background functions $\psi$ and $\rho$. We now want to plug these solutions into Eq. (\ref{eq:S2}) to obtain a partially on-shell action. After integrating by parts and using the background equations of motion, the resulting low-energy action is found to be
\begin{align} \label{eq:Salmost}
S^{(2)}=&-\frac{\sqrt{2}}{2q^2\mu}\int d^{D+1}x\bigg[-4q^2\,\frac{\psi\rho\rho'}{u^{D-1}}\,\partial_0\pi_B\partial_0\pi-2q^2\,\frac{\rho^2\psi'}{u^{D-1}}\,\partial_0\pi_B\partial_0\pi \notag \\
&+2q^2\,\frac{\psi\rho^2}{u^D}\,\partial_i\pi_B\partial_i\pi-\frac{\psi''}{u^{D-3}}\,\partial_0\pi_B\partial_0\pi'+\frac{(\psi/u)'}{u^{D-3}}\,\partial_i\pi_B\partial_i\pi'\bigg] +S^{(2)}_\text{c.t.}\,.
\end{align}
Interestingly, the above action does not depend on the scalar field potential. All the boundary terms up to this point vanish because they are subleading as $u\to0$ and/or of higher order in $\epsilon$.

Moreover, we will now show in general that the counterterm $S^{(2)}_\text{c.t.}$ does not contribute to the quadratic phonon action either. The general expression for this term, at lowest order in boundary derivatives, is
\begin{align}
S_\text{c.t.}=\int d^Dx\,\sqrt{-\gamma}\,\mathcal{L}_\text{c.t.}(\Phi,D_M\Phi,F_{MN}),
\end{align}
where $\mathcal{L}_\text{c.t.}$ is an analytic function of gauge and diffeomorphism invariant combinations of its arguments and their derivatives. When expanded in small fluctuations of the fields, it will have the schematic form
\begin{align}
\mathcal{L}_\text{c.t.}\sim \mathcal{L}_\text{c.t.}^{(0)}+\mathcal{L}_\text{c.t.}^{(1)}\delta\Psi+\mathcal{L}_\text{c.t.}^{(2)}\delta\Psi^2+\dots
\end{align}
where we have collectively denoted the fluctuations by $\delta\Psi$. Moreover, since the background bulk action is finite when $\rho_{(1)}=0$, the $\mathcal{L}_\text{c.t.}^{(n)}$ terms must be either finite or log divergent in the $u \to 0$ limit---see e.g.\ \cite{Papadimitriou:2010as}. This ensures that $\mathcal{L}_\text{c.t.}$ will vanish if all possible fluctuations $\delta\Psi$ scale as some positive powers of $u$. We will now argue that this is indeed the case.

To this end, let us perform a power counting in $u\sim0$. Note first that pairs of indices can be contracted with the metric, $g^{MN}\sim u^2$, while a single radial index could also be contracted with the vector normal to the boundary $n^M\sim u$. It follows that each free covariant index carries a factor of $u$. But then, from Eqs.~\eqref{falloffs}, \eqref{eq:ai_sol} and \eqref{eq:sigma_a0_sol}, it is easy to derive the following scalings
\begin{subequations}
\begin{gather}
\Phi\sim u^\Delta,\quad D_u\Phi \sim u^\Delta,\\
 D_\mu\Phi\sim u^{\Delta+1}, \quad F_{u\mu}\sim u^{D-1},
\end{gather}
\end{subequations}
while $F_{\mu\nu}$ is of higher order in boundary derivatives. The above behaviors are true both for the background and for the fluctuations. Indeed, based on the solutions in Eq.~\eqref{eq:sigma_a0_sol}, we see that $\sigma\sim u\rho^\prime\sim\rho$, $f_{u0}\sim \psi^{\prime\prime} \sim (\psi/u)^\prime \sim \bar F_{u0}$ and $F_{ui}=f_{ui}$.

A generic gauge and diffeomorphism invariant term will then look schematically like
\begin{align} \label{powercounting}
\sqrt{-\gamma}\,\Phi^n (D_M\Phi)^m (F_{MN})^\ell\sim u^{-D+(n+m)\Delta+\ell(D-1)},
\end{align}
where the covariant indices should be thought of as contracted either with $g^{MN}$ or with $n^M$. Gauge invariance requires each operator to be neutral, and hence to include as many factors of $\Phi$ as of $\Phi^{*}$. This implies that $n+m$ must be an even nonnegative integer and therefore Eq.~\eqref{powercounting} always goes to zero as a positive power of $u$. To see this explicitly, note that given the definition of $\Delta$ presented below Eqs.~\eqref{falloff}, it must be that $\Delta>D/2$. Thus if $n+m\geq2$ one has
\begin{align}
-D+(n+m)\Delta+\ell(D-1)&>(n+m-2)\Delta+\ell(D-1)\geq0\,. \notag
\end{align}
If instead $n+m=0$, diffeomorphism invariance implies $\ell\geq 2$ and hence $-D+\ell(D-1)>0$, again for $D>2$. This ensures that the counterterm does not contribute to the on-shell quadratic action.

\vspace{1em}

Now, recalling the near boundary behavior of $\psi(u)$ given in Eq.~\eqref{asymppsi}, integrating by parts Eq.~\eqref{eq:Salmost} and using the equations of motion for the background gauge field, the quadratic action reduces to a purely boundary term:
\begin{align} \label{quadratic S}
S^{(2)}=\frac{\varepsilon(D-1)(D-2)}{2 q^2\mu}\!\int \!d^Dx\!\left[\dot \pi_B^2-\frac{\partial_i \pi_B \partial^i \pi_B}{(D-1)}  \right] \! .  
\end{align}
This is indeed the right action for free phonons in a conformal superfluid, with $c_s^2=1/(D-1)$. The overall coefficient in (\ref{quadratic S}) is positive for $D>2$ because $\varepsilon / \mu >0$ with our conventions, thus ensuring that $\pi_B$ is not ghost-like. Our argument does not apply to the $D=2$ case since the asymptotic behavior of the fields is not regular anymore. However in this case we do not expect Goldstone modes because of Coleman's theorem~\cite{Coleman:1973ci,Anninos:2010sq}. Finally, it is interesting to note how the boundary Goldstone boson $\pi_B$ turns out to be the Wilson line of the radial component of the bulk gauge field---see Eq. (\ref{eq:pi_sol}). This is the same result found in~\cite{deBoer:2015ija} for ordinary fluids and in~\cite{Nickel:2010pr} for a Maxwell-Einstein theory, and it also resembles the results found for pions in models of holographic QCD~\cite{Sakai:2004cn}.\\

\bigskip

\section{Discussion}

Holographic superfluids at zero temperature provide a simple context in which to illustrate the techniques developed in~\cite{Nickel:2010pr,deBoer:2015ija,Crossley:2015tka}. At the same time, in this paper we further extended the reach of these methods and showed that an explicit expression for the background fields is not needed in order to derive the boundary action for the Goldstones. This paves the way for the application of these methods to other condensed matter systems~\cite{Nicolis:2015sra}.

It would be interesting to see if the boundary Goldstone action can be derived, using the same procedure, when including the backreaction of the fields on the geometry. In particular, the value of the superfluid speed of sound could in general be model dependent, i.e.\ depend on the specific scalar field potential, since the boundary terms arising from integrating the action by parts may in principle give contributions that differ from the ones found in the probe approximation. One compelling case to study is the W-shaped potential analyzed in~\cite{Gubser:2008wz,Gubser:2008pf,Gubser:2009cg}. There the bulk geometry is found to exhibit (for a charge $q$ greater than some critical value) a domain wall that interpolates between an IR and a UV AdS regions with different curvature radii. Thus we expect that, in this setting, the asymptotic behavior of the fields be essentially equivalent to those in the absence of backreaction, and hence that our final results should not change. This makes this model particularly appealing for testing our analytical method beyond the probe limit.

As another further development of the techniques we have presented, it would be interesting to find a holographic derivation of the effective action for superfluids at finite temperatures~\cite{Nicolis:2011cs}. Previous holographic models of finite temperature superfluids were found not to reproduce Landau's prediction for the relation between second and first sound at low temperature~\cite{Herzog:2009md}. As pointed out by the same authors, this might be due to the presence of additional degrees of freedom. Our approach might be able to explicitly isolate these degrees of freedom and hence shed more light on this issue. We leave these interesting research questions for future work.
\\
\vspace{1em}

\begin{acknowledgments}
We are very grateful to R.\ Krichevsky for collaboration in the early stages of this project, and to A.\ Nicolis for extensive conversations and feedback on the manuscript. We would also like to thank C.\ Asplund, F.\ Denef, W.\ Goldberger, S.\ Hartnoll, K.\ Jensen, R.\ Monten, I.\ Papadimitriou, R.\ Rattazzi and especially C.\ Herzog, C.\ Toldo and A.\ Yarom for extremely helpful conversations and comments. This work was supported in part by the United States Department of Energy under contracts DE-FG02-11ER41743 and DE-FG02-92-ER40699.
\end{acknowledgments}

\bibliographystyle{apsrev4-1}
\bibliography{biblio}

\begin{thebibliography}{29}%
\makeatletter
\providecommand \@ifxundefined [1]{%
 \@ifx{#1\undefined}
}%
\providecommand \@ifnum [1]{%
 \ifnum #1\expandafter \@firstoftwo
 \else \expandafter \@secondoftwo
 \fi
}%
\providecommand \@ifx [1]{%
 \ifx #1\expandafter \@firstoftwo
 \else \expandafter \@secondoftwo
 \fi
}%
\providecommand \natexlab [1]{#1}%
\providecommand \enquote  [1]{``#1''}%
\providecommand \bibnamefont  [1]{#1}%
\providecommand \bibfnamefont [1]{#1}%
\providecommand \citenamefont [1]{#1}%
\providecommand \href@noop [0]{\@secondoftwo}%
\providecommand \href [0]{\begingroup \@sanitize@url \@href}%
\providecommand \@href[1]{\@@startlink{#1}\@@href}%
\providecommand \@@href[1]{\endgroup#1\@@endlink}%
\providecommand \@sanitize@url [0]{\catcode `\\12\catcode `\$12\catcode
  `\&12\catcode `\#12\catcode `\^12\catcode `\_12\catcode `\%12\relax}%
\providecommand \@@startlink[1]{}%
\providecommand \@@endlink[0]{}%
\providecommand \url  [0]{\begingroup\@sanitize@url \@url }%
\providecommand \@url [1]{\endgroup\@href {#1}{\urlprefix }}%
\providecommand \urlprefix  [0]{URL }%
\providecommand \Eprint [0]{\href }%
\providecommand \doibase [0]{http://dx.doi.org/}%
\providecommand \selectlanguage [0]{\@gobble}%
\providecommand \bibinfo  [0]{\@secondoftwo}%
\providecommand \bibfield  [0]{\@secondoftwo}%
\providecommand \translation [1]{[#1]}%
\providecommand \BibitemOpen [0]{}%
\providecommand \bibitemStop [0]{}%
\providecommand \bibitemNoStop [0]{.\EOS\space}%
\providecommand \EOS [0]{\spacefactor3000\relax}%
\providecommand \BibitemShut  [1]{\csname bibitem#1\endcsname}%
\let\auto@bib@innerbib\@empty
\bibitem [{\citenamefont {Maldacena}(1999)}]{Maldacena:1997re}%
  \BibitemOpen
  \bibfield  {author} {\bibinfo {author} {\bibfnamefont {J.~M.}\ \bibnamefont
  {Maldacena}},\ }\href {\doibase 10.1023/A:1026654312961} {\bibfield
  {journal} {\bibinfo  {journal} {Int. J. Theor. Phys.}\ }\textbf {\bibinfo
  {volume} {38}},\ \bibinfo {pages} {1113} (\bibinfo {year} {1999})},\ \bibinfo
  {note} {[Adv. Theor. Math. Phys.2,231(1998)]},\ \Eprint
  {http://arxiv.org/abs/hep-th/9711200} {arXiv:hep-th/9711200 [hep-th]}
  \BibitemShut {NoStop}%
\bibitem [{\citenamefont {Gubser}\ \emph {et~al.}(1998)\citenamefont {Gubser},
  \citenamefont {Klebanov},\ and\ \citenamefont {Polyakov}}]{Gubser:1998bc}%
  \BibitemOpen
  \bibfield  {author} {\bibinfo {author} {\bibfnamefont {S.~S.}\ \bibnamefont
  {Gubser}}, \bibinfo {author} {\bibfnamefont {I.~R.}\ \bibnamefont
  {Klebanov}}, \ and\ \bibinfo {author} {\bibfnamefont {A.~M.}\ \bibnamefont
  {Polyakov}},\ }\href {\doibase 10.1016/S0370-2693(98)00377-3} {\bibfield
  {journal} {\bibinfo  {journal} {Phys. Lett.}\ }\textbf {\bibinfo {volume}
  {B428}},\ \bibinfo {pages} {105} (\bibinfo {year} {1998})},\ \Eprint
  {http://arxiv.org/abs/hep-th/9802109} {arXiv:hep-th/9802109 [hep-th]}
  \BibitemShut {NoStop}%
\bibitem [{\citenamefont {Witten}(1998)}]{Witten:1998qj}%
  \BibitemOpen
  \bibfield  {author} {\bibinfo {author} {\bibfnamefont {E.}~\bibnamefont
  {Witten}},\ }\href@noop {} {\bibfield  {journal} {\bibinfo  {journal} {Adv.
  Theor. Math. Phys.}\ }\textbf {\bibinfo {volume} {2}},\ \bibinfo {pages}
  {253} (\bibinfo {year} {1998})},\ \Eprint
  {http://arxiv.org/abs/hep-th/9802150} {arXiv:hep-th/9802150 [hep-th]}
  \BibitemShut {NoStop}%
\bibitem [{\citenamefont {Hartnoll}(2009)}]{Hartnoll:2009sz}%
  \BibitemOpen
  \bibfield  {author} {\bibinfo {author} {\bibfnamefont {S.~A.}\ \bibnamefont
  {Hartnoll}},\ }\href@noop {} {\bibfield  {journal} {\bibinfo  {journal}
  {Class. Quant. Grav.}\ }\textbf {\bibinfo {volume} {26}},\ \bibinfo {pages}
  {224002} (\bibinfo {year} {2009})},\ \Eprint {http://arxiv.org/abs/0903.3246}
  {arXiv:0903.3246 [hep-th]} \BibitemShut {NoStop}%
\bibitem [{\citenamefont {Herzog}(2009)}]{Herzog:2009xv}%
  \BibitemOpen
  \bibfield  {author} {\bibinfo {author} {\bibfnamefont {C.~P.}\ \bibnamefont
  {Herzog}},\ }\href@noop {} {\bibfield  {journal} {\bibinfo  {journal} {J.
  Phys.}\ }\textbf {\bibinfo {volume} {A42}},\ \bibinfo {pages} {343001}
  (\bibinfo {year} {2009})},\ \Eprint {http://arxiv.org/abs/0904.1975}
  {arXiv:0904.1975 [hep-th]} \BibitemShut {NoStop}%
\bibitem [{\citenamefont {Zaanen}\ \emph {et~al.}(2015)\citenamefont {Zaanen},
  \citenamefont {Sun}, \citenamefont {Liu},\ and\ \citenamefont
  {Schalm}}]{Zaanen:2015oix}%
  \BibitemOpen
  \bibfield  {author} {\bibinfo {author} {\bibfnamefont {J.}~\bibnamefont
  {Zaanen}}, \bibinfo {author} {\bibfnamefont {Y.-W.}\ \bibnamefont {Sun}},
  \bibinfo {author} {\bibfnamefont {Y.}~\bibnamefont {Liu}}, \ and\ \bibinfo
  {author} {\bibfnamefont {K.}~\bibnamefont {Schalm}},\ }\href
  {http://www.cambridge.org/mw/academic/subjects/physics/condensed-matter-physics-nanoscience-and-mesoscopic-physics/holographic-duality-condensed-matter-physics?format=HB}
  {\emph {\bibinfo {title} {{Holographic Duality in Condensed Matter
  Physics}}}}\ (\bibinfo  {publisher} {Cambridge Univ. Press},\ \bibinfo {year}
  {2015})\BibitemShut {NoStop}%
\bibitem [{\citenamefont {Nicolis}\ and\ \citenamefont
  {Piazza}(2012)}]{Nicolis:2011pv}%
  \BibitemOpen
  \bibfield  {author} {\bibinfo {author} {\bibfnamefont {A.}~\bibnamefont
  {Nicolis}}\ and\ \bibinfo {author} {\bibfnamefont {F.}~\bibnamefont
  {Piazza}},\ }\href {\doibase 10.1007/JHEP06(2012)025} {\bibfield  {journal}
  {\bibinfo  {journal} {JHEP}\ }\textbf {\bibinfo {volume} {06}},\ \bibinfo
  {pages} {025} (\bibinfo {year} {2012})},\ \Eprint
  {http://arxiv.org/abs/1112.5174} {arXiv:1112.5174 [hep-th]} \BibitemShut
  {NoStop}%
\bibitem [{\citenamefont {Nicolis}\ \emph {et~al.}(2014)\citenamefont
  {Nicolis}, \citenamefont {Penco},\ and\ \citenamefont
  {Rosen}}]{Nicolis:2013lma}%
  \BibitemOpen
  \bibfield  {author} {\bibinfo {author} {\bibfnamefont {A.}~\bibnamefont
  {Nicolis}}, \bibinfo {author} {\bibfnamefont {R.}~\bibnamefont {Penco}}, \
  and\ \bibinfo {author} {\bibfnamefont {R.~A.}\ \bibnamefont {Rosen}},\ }\href
  {\doibase 10.1103/PhysRevD.89.045002} {\bibfield  {journal} {\bibinfo
  {journal} {Phys.Rev.}\ }\textbf {\bibinfo {volume} {D89}},\ \bibinfo {pages}
  {045002} (\bibinfo {year} {2014})},\ \Eprint {http://arxiv.org/abs/1307.0517}
  {arXiv:1307.0517 [hep-th]} \BibitemShut {NoStop}%
\bibitem [{\citenamefont {Son}(2002)}]{Son:2002zn}%
  \BibitemOpen
  \bibfield  {author} {\bibinfo {author} {\bibfnamefont {D.~T.}\ \bibnamefont
  {Son}},\ }\href@noop {} {\  (\bibinfo {year} {2002})},\ \Eprint
  {http://arxiv.org/abs/hep-ph/0204199} {arXiv:hep-ph/0204199 [hep-ph]}
  \BibitemShut {NoStop}%
\bibitem [{\citenamefont {Nicolis}(2011)}]{Nicolis:2011cs}%
  \BibitemOpen
  \bibfield  {author} {\bibinfo {author} {\bibfnamefont {A.}~\bibnamefont
  {Nicolis}},\ }\href@noop {} {\  (\bibinfo {year} {2011})},\ \Eprint
  {http://arxiv.org/abs/1108.2513} {arXiv:1108.2513 [hep-th]} \BibitemShut
  {NoStop}%
\bibitem [{\citenamefont {Herzog}\ \emph {et~al.}(2009)\citenamefont {Herzog},
  \citenamefont {Kovtun},\ and\ \citenamefont {Son}}]{Herzog:2008he}%
  \BibitemOpen
  \bibfield  {author} {\bibinfo {author} {\bibfnamefont {C.~P.}\ \bibnamefont
  {Herzog}}, \bibinfo {author} {\bibfnamefont {P.~K.}\ \bibnamefont {Kovtun}},
  \ and\ \bibinfo {author} {\bibfnamefont {D.~T.}\ \bibnamefont {Son}},\ }\href
  {\doibase 10.1103/PhysRevD.79.066002} {\bibfield  {journal} {\bibinfo
  {journal} {Phys. Rev.}\ }\textbf {\bibinfo {volume} {D79}},\ \bibinfo {pages}
  {066002} (\bibinfo {year} {2009})},\ \Eprint {http://arxiv.org/abs/0809.4870}
  {arXiv:0809.4870 [hep-th]} \BibitemShut {NoStop}%
\bibitem [{\citenamefont {Hartnoll}\ \emph
  {et~al.}(2008{\natexlab{a}})\citenamefont {Hartnoll}, \citenamefont
  {Herzog},\ and\ \citenamefont {Horowitz}}]{Hartnoll:2008vx}%
  \BibitemOpen
  \bibfield  {author} {\bibinfo {author} {\bibfnamefont {S.~A.}\ \bibnamefont
  {Hartnoll}}, \bibinfo {author} {\bibfnamefont {C.~P.}\ \bibnamefont
  {Herzog}}, \ and\ \bibinfo {author} {\bibfnamefont {G.~T.}\ \bibnamefont
  {Horowitz}},\ }\href {\doibase 10.1103/PhysRevLett.101.031601} {\bibfield
  {journal} {\bibinfo  {journal} {Phys. Rev. Lett.}\ }\textbf {\bibinfo
  {volume} {101}},\ \bibinfo {pages} {031601} (\bibinfo {year}
  {2008}{\natexlab{a}})},\ \Eprint {http://arxiv.org/abs/0803.3295}
  {arXiv:0803.3295 [hep-th]} \BibitemShut {NoStop}%
\bibitem [{\citenamefont {Hartnoll}\ \emph
  {et~al.}(2008{\natexlab{b}})\citenamefont {Hartnoll}, \citenamefont
  {Herzog},\ and\ \citenamefont {Horowitz}}]{Hartnoll:2008kx}%
  \BibitemOpen
  \bibfield  {author} {\bibinfo {author} {\bibfnamefont {S.~A.}\ \bibnamefont
  {Hartnoll}}, \bibinfo {author} {\bibfnamefont {C.~P.}\ \bibnamefont
  {Herzog}}, \ and\ \bibinfo {author} {\bibfnamefont {G.~T.}\ \bibnamefont
  {Horowitz}},\ }\href {\doibase 10.1088/1126-6708/2008/12/015} {\bibfield
  {journal} {\bibinfo  {journal} {JHEP}\ }\textbf {\bibinfo {volume} {12}},\
  \bibinfo {pages} {015} (\bibinfo {year} {2008}{\natexlab{b}})},\ \Eprint
  {http://arxiv.org/abs/0810.1563} {arXiv:0810.1563 [hep-th]} \BibitemShut
  {NoStop}%
\bibitem [{\citenamefont {Herzog}\ and\ \citenamefont
  {Yarom}(2009)}]{Herzog:2009md}%
  \BibitemOpen
  \bibfield  {author} {\bibinfo {author} {\bibfnamefont {C.~P.}\ \bibnamefont
  {Herzog}}\ and\ \bibinfo {author} {\bibfnamefont {A.}~\bibnamefont {Yarom}},\
  }\href {\doibase 10.1103/PhysRevD.80.106002} {\bibfield  {journal} {\bibinfo
  {journal} {Phys. Rev.}\ }\textbf {\bibinfo {volume} {D80}},\ \bibinfo {pages}
  {106002} (\bibinfo {year} {2009})},\ \Eprint {http://arxiv.org/abs/0906.4810}
  {arXiv:0906.4810 [hep-th]} \BibitemShut {NoStop}%
\bibitem [{\citenamefont {Yarom}(2009)}]{Yarom:2009uq}%
  \BibitemOpen
  \bibfield  {author} {\bibinfo {author} {\bibfnamefont {A.}~\bibnamefont
  {Yarom}},\ }\href {\doibase 10.1088/1126-6708/2009/07/070} {\bibfield
  {journal} {\bibinfo  {journal} {JHEP}\ }\textbf {\bibinfo {volume} {07}},\
  \bibinfo {pages} {070} (\bibinfo {year} {2009})},\ \Eprint
  {http://arxiv.org/abs/0903.1353} {arXiv:0903.1353 [hep-th]} \BibitemShut
  {NoStop}%
\bibitem [{\citenamefont {Nickel}\ and\ \citenamefont
  {Son}(2011)}]{Nickel:2010pr}%
  \BibitemOpen
  \bibfield  {author} {\bibinfo {author} {\bibfnamefont {D.}~\bibnamefont
  {Nickel}}\ and\ \bibinfo {author} {\bibfnamefont {D.~T.}\ \bibnamefont
  {Son}},\ }\href {\doibase 10.1088/1367-2630/13/7/075010} {\bibfield
  {journal} {\bibinfo  {journal} {New J. Phys.}\ }\textbf {\bibinfo {volume}
  {13}},\ \bibinfo {pages} {075010} (\bibinfo {year} {2011})},\ \Eprint
  {http://arxiv.org/abs/1009.3094} {arXiv:1009.3094 [hep-th]} \BibitemShut
  {NoStop}%
\bibitem [{\citenamefont {de~Boer}\ \emph {et~al.}(2015)\citenamefont
  {de~Boer}, \citenamefont {Heller},\ and\ \citenamefont
  {Pinzani-Fokeeva}}]{deBoer:2015ija}%
  \BibitemOpen
  \bibfield  {author} {\bibinfo {author} {\bibfnamefont {J.}~\bibnamefont
  {de~Boer}}, \bibinfo {author} {\bibfnamefont {M.~P.}\ \bibnamefont {Heller}},
  \ and\ \bibinfo {author} {\bibfnamefont {N.}~\bibnamefont
  {Pinzani-Fokeeva}},\ }\href {\doibase 10.1007/JHEP08(2015)086} {\bibfield
  {journal} {\bibinfo  {journal} {JHEP}\ }\textbf {\bibinfo {volume} {08}},\
  \bibinfo {pages} {086} (\bibinfo {year} {2015})},\ \Eprint
  {http://arxiv.org/abs/1504.07616} {arXiv:1504.07616 [hep-th]} \BibitemShut
  {NoStop}%
\bibitem [{\citenamefont {Crossley}\ \emph {et~al.}(2016)\citenamefont
  {Crossley}, \citenamefont {Glorioso}, \citenamefont {Liu},\ and\
  \citenamefont {Wang}}]{Crossley:2015tka}%
  \BibitemOpen
  \bibfield  {author} {\bibinfo {author} {\bibfnamefont {M.}~\bibnamefont
  {Crossley}}, \bibinfo {author} {\bibfnamefont {P.}~\bibnamefont {Glorioso}},
  \bibinfo {author} {\bibfnamefont {H.}~\bibnamefont {Liu}}, \ and\ \bibinfo
  {author} {\bibfnamefont {Y.}~\bibnamefont {Wang}},\ }\href {\doibase
  10.1007/JHEP02(2016)124} {\bibfield  {journal} {\bibinfo  {journal} {JHEP}\
  }\textbf {\bibinfo {volume} {02}},\ \bibinfo {pages} {124} (\bibinfo {year}
  {2016})},\ \Eprint {http://arxiv.org/abs/1504.07611} {arXiv:1504.07611
  [hep-th]} \BibitemShut {NoStop}%
\bibitem [{\citenamefont {Argurio}\ \emph {et~al.}(2015)\citenamefont
  {Argurio}, \citenamefont {Marzolla}, \citenamefont {Mezzalira},\ and\
  \citenamefont {Naegels}}]{Argurio:2015via}%
  \BibitemOpen
  \bibfield  {author} {\bibinfo {author} {\bibfnamefont {R.}~\bibnamefont
  {Argurio}}, \bibinfo {author} {\bibfnamefont {A.}~\bibnamefont {Marzolla}},
  \bibinfo {author} {\bibfnamefont {A.}~\bibnamefont {Mezzalira}}, \ and\
  \bibinfo {author} {\bibfnamefont {D.}~\bibnamefont {Naegels}},\ }\href
  {\doibase 10.1103/PhysRevD.92.066009} {\bibfield  {journal} {\bibinfo
  {journal} {Phys. Rev.}\ }\textbf {\bibinfo {volume} {D92}},\ \bibinfo {pages}
  {066009} (\bibinfo {year} {2015})},\ \Eprint
  {http://arxiv.org/abs/1507.00211} {arXiv:1507.00211 [hep-th]} \BibitemShut
  {NoStop}%
\bibitem [{\citenamefont {Argurio}\ \emph {et~al.}(2016)\citenamefont
  {Argurio}, \citenamefont {Marzolla}, \citenamefont {Mezzalira},\ and\
  \citenamefont {Musso}}]{Argurio:2015wgr}%
  \BibitemOpen
  \bibfield  {author} {\bibinfo {author} {\bibfnamefont {R.}~\bibnamefont
  {Argurio}}, \bibinfo {author} {\bibfnamefont {A.}~\bibnamefont {Marzolla}},
  \bibinfo {author} {\bibfnamefont {A.}~\bibnamefont {Mezzalira}}, \ and\
  \bibinfo {author} {\bibfnamefont {D.}~\bibnamefont {Musso}},\ }\href
  {\doibase 10.1007/JHEP03(2016)012} {\bibfield  {journal} {\bibinfo  {journal}
  {JHEP}\ }\textbf {\bibinfo {volume} {03}},\ \bibinfo {pages} {012} (\bibinfo
  {year} {2016})},\ \Eprint {http://arxiv.org/abs/1512.03750} {arXiv:1512.03750
  [hep-th]} \BibitemShut {NoStop}%
\bibitem [{\citenamefont {Horowitz}\ and\ \citenamefont
  {Roberts}(2009)}]{Horowitz:2009ij}%
  \BibitemOpen
  \bibfield  {author} {\bibinfo {author} {\bibfnamefont {G.~T.}\ \bibnamefont
  {Horowitz}}\ and\ \bibinfo {author} {\bibfnamefont {M.~M.}\ \bibnamefont
  {Roberts}},\ }\href {\doibase 10.1088/1126-6708/2009/11/015} {\bibfield
  {journal} {\bibinfo  {journal} {JHEP}\ }\textbf {\bibinfo {volume} {11}},\
  \bibinfo {pages} {015} (\bibinfo {year} {2009})},\ \Eprint
  {http://arxiv.org/abs/0908.3677} {arXiv:0908.3677 [hep-th]} \BibitemShut
  {NoStop}%
\bibitem [{\citenamefont {Gubser}\ and\ \citenamefont
  {Nellore}(2009{\natexlab{a}})}]{Gubser:2009cg}%
  \BibitemOpen
  \bibfield  {author} {\bibinfo {author} {\bibfnamefont {S.~S.}\ \bibnamefont
  {Gubser}}\ and\ \bibinfo {author} {\bibfnamefont {A.}~\bibnamefont
  {Nellore}},\ }\href {\doibase 10.1103/PhysRevD.80.105007} {\bibfield
  {journal} {\bibinfo  {journal} {Phys. Rev.}\ }\textbf {\bibinfo {volume}
  {D80}},\ \bibinfo {pages} {105007} (\bibinfo {year} {2009}{\natexlab{a}})},\
  \Eprint {http://arxiv.org/abs/0908.1972} {arXiv:0908.1972 [hep-th]}
  \BibitemShut {NoStop}%
\bibitem [{\citenamefont {Gubser}\ and\ \citenamefont
  {Rocha}(2009)}]{Gubser:2008wz}%
  \BibitemOpen
  \bibfield  {author} {\bibinfo {author} {\bibfnamefont {S.~S.}\ \bibnamefont
  {Gubser}}\ and\ \bibinfo {author} {\bibfnamefont {F.~D.}\ \bibnamefont
  {Rocha}},\ }\href {\doibase 10.1103/PhysRevLett.102.061601} {\bibfield
  {journal} {\bibinfo  {journal} {Phys. Rev. Lett.}\ }\textbf {\bibinfo
  {volume} {102}},\ \bibinfo {pages} {061601} (\bibinfo {year} {2009})},\
  \Eprint {http://arxiv.org/abs/0807.1737} {arXiv:0807.1737 [hep-th]}
  \BibitemShut {NoStop}%
\bibitem [{\citenamefont {Papadimitriou}(2010)}]{Papadimitriou:2010as}%
  \BibitemOpen
  \bibfield  {author} {\bibinfo {author} {\bibfnamefont {I.}~\bibnamefont
  {Papadimitriou}},\ }\href {\doibase 10.1007/JHEP11(2010)014} {\bibfield
  {journal} {\bibinfo  {journal} {JHEP}\ }\textbf {\bibinfo {volume} {11}},\
  \bibinfo {pages} {014} (\bibinfo {year} {2010})},\ \Eprint
  {http://arxiv.org/abs/1007.4592} {arXiv:1007.4592 [hep-th]} \BibitemShut
  {NoStop}%
\bibitem [{\citenamefont {Coleman}(1973)}]{Coleman:1973ci}%
  \BibitemOpen
  \bibfield  {author} {\bibinfo {author} {\bibfnamefont {S.~R.}\ \bibnamefont
  {Coleman}},\ }\href {\doibase 10.1007/BF01646487} {\bibfield  {journal}
  {\bibinfo  {journal} {Commun.Math.Phys.}\ }\textbf {\bibinfo {volume} {31}},\
  \bibinfo {pages} {259} (\bibinfo {year} {1973})}\BibitemShut {NoStop}%
\bibitem [{\citenamefont {Anninos}\ \emph {et~al.}(2010)\citenamefont
  {Anninos}, \citenamefont {Hartnoll},\ and\ \citenamefont
  {Iqbal}}]{Anninos:2010sq}%
  \BibitemOpen
  \bibfield  {author} {\bibinfo {author} {\bibfnamefont {D.}~\bibnamefont
  {Anninos}}, \bibinfo {author} {\bibfnamefont {S.~A.}\ \bibnamefont
  {Hartnoll}}, \ and\ \bibinfo {author} {\bibfnamefont {N.}~\bibnamefont
  {Iqbal}},\ }\href {\doibase 10.1103/PhysRevD.82.066008} {\bibfield  {journal}
  {\bibinfo  {journal} {Phys. Rev.}\ }\textbf {\bibinfo {volume} {D82}},\
  \bibinfo {pages} {066008} (\bibinfo {year} {2010})},\ \Eprint
  {http://arxiv.org/abs/1005.1973} {arXiv:1005.1973 [hep-th]} \BibitemShut
  {NoStop}%
\bibitem [{\citenamefont {Sakai}\ and\ \citenamefont
  {Sugimoto}(2005)}]{Sakai:2004cn}%
  \BibitemOpen
  \bibfield  {author} {\bibinfo {author} {\bibfnamefont {T.}~\bibnamefont
  {Sakai}}\ and\ \bibinfo {author} {\bibfnamefont {S.}~\bibnamefont
  {Sugimoto}},\ }\href {\doibase 10.1143/PTP.113.843} {\bibfield  {journal}
  {\bibinfo  {journal} {Prog. Theor. Phys.}\ }\textbf {\bibinfo {volume}
  {113}},\ \bibinfo {pages} {843} (\bibinfo {year} {2005})},\ \Eprint
  {http://arxiv.org/abs/hep-th/0412141} {arXiv:hep-th/0412141 [hep-th]}
  \BibitemShut {NoStop}%
\bibitem [{\citenamefont {Nicolis}\ \emph {et~al.}(2015)\citenamefont
  {Nicolis}, \citenamefont {Penco}, \citenamefont {Piazza},\ and\ \citenamefont
  {Rattazzi}}]{Nicolis:2015sra}%
  \BibitemOpen
  \bibfield  {author} {\bibinfo {author} {\bibfnamefont {A.}~\bibnamefont
  {Nicolis}}, \bibinfo {author} {\bibfnamefont {R.}~\bibnamefont {Penco}},
  \bibinfo {author} {\bibfnamefont {F.}~\bibnamefont {Piazza}}, \ and\ \bibinfo
  {author} {\bibfnamefont {R.}~\bibnamefont {Rattazzi}},\ }\href {\doibase
  10.1007/JHEP06(2015)155} {\bibfield  {journal} {\bibinfo  {journal} {JHEP}\
  }\textbf {\bibinfo {volume} {06}},\ \bibinfo {pages} {155} (\bibinfo {year}
  {2015})},\ \Eprint {http://arxiv.org/abs/1501.03845} {arXiv:1501.03845
  [hep-th]} \BibitemShut {NoStop}%
\bibitem [{\citenamefont {Gubser}\ and\ \citenamefont
  {Nellore}(2009{\natexlab{b}})}]{Gubser:2008pf}%
  \BibitemOpen
  \bibfield  {author} {\bibinfo {author} {\bibfnamefont {S.~S.}\ \bibnamefont
  {Gubser}}\ and\ \bibinfo {author} {\bibfnamefont {A.}~\bibnamefont
  {Nellore}},\ }\href {\doibase 10.1088/1126-6708/2009/04/008} {\bibfield
  {journal} {\bibinfo  {journal} {JHEP}\ }\textbf {\bibinfo {volume} {04}},\
  \bibinfo {pages} {008} (\bibinfo {year} {2009}{\natexlab{b}})},\ \Eprint
  {http://arxiv.org/abs/0810.4554} {arXiv:0810.4554 [hep-th]} \BibitemShut
  {NoStop}%
\end{thebibliography}%

\end{document}